# Modulation-assisted time-delay interferometric ranging for LISA

**Jean-Baptiste Bayle**[1,7,*], **Martin Staab**[2,3,4,7], **Samuel P Francis**[5], **Emily Rose Rees**[6], **Robert Spero**[5] **and Gerhard Heinzel**[2,3]

[1] University of Glasgow, Glasgow G12 8QQ, United Kingdom
[2] Max-Planck-Institut für Gravitationsphysik (Albert-Einstein-Institut), Callinstraße 38, 30167 Hannover, Germany
[3] Leibniz Universität Hannover, D-30167 Hanover, Germany
[4] SYRTE, Observatoire de Paris, Université PSL, CNRS, Sorbonne Université, LNE, Paris, France
[5] Jet Propulsion Laboratory, California Institute of Technology, 4800 Oak Grove Drive, Pasadena, CA 91109, United States of America
[6] Centre for Gravitational Astrophysics (CGA), Research School of Physics, The Australian National University, Canberra ACT 2601, Australia

E-mail: j2b.bayle@gmail.com

## Abstract

Laser Interferometer Space Antenna (LISA) represents the next frontier in gravitational-wave (GW) astronomy, targeting the detection of millihertz gravitational signals. Central to LISA's operation is the nanosecond-precision estimation of the light travel times (LTTs) between its constituent spacecraft. Precise LTT estimates are critical for suppressing dominant laser noise with time-delay interferometry (TDI) and ensuring the required sensitivity to GW signals. The baseline method is to modulate a pseudorandom noise (PRN) code on the laser beams exchanged between the spacecraft. Time-delay interferometric ranging (TDIR) was proposed as a simpler alternative LTT estimation method. TDIR LTT estimates are chosen to minimize the TDI residual noise over the full LISA frequency band. TDIR can be used in case of PRN failure, or to calibrate the biases of the PRN method. In this study, we introduce modulation-assisted TDIR (MATDIR), an enhanced variant of TDIR that significantly improves LTT estimation precision and resilience. MATDIR achieves this by modulating the laser phase at specific frequencies, close to 1 Hz, thereby artificially elevating the laser phase content relative to secondary, unsuppressed noises. This modulation strategy not only enhances the signal-to-noise ratio for TDIR but also mitigates the impact of GW signals and instrumental artifacts, enabling more reliable LTT estimates with reduced integration times. We develop the theoretical framework of MATDIR, incorporating the full constellation of three spacecraft, laser locking, and multiple Michelson TDI combinations. Analytical predictions, confirmed by numerical simulations, indicate that MATDIR can achieve LTT estimates comparable to the 1 m-rms at $f_s = 4$ Hz of the PRN-based baseline method. We therefore suggest that the possibility to modulate lasers is added to the laser system requirements of LISA.

---

[7] These authors contributed equally to this work.
[*] Author to whom any correspondence should be addressed.

## 1. Introduction

Following the many observations of ground-based gravitational-wave (GW) detectors [1, 2], the European Space Agency (ESA) has recently adopted Laser Interferometer Space Antenna (LISA) as the next large mission. Its launch date is scheduled for 2035. LISA is a spaceborne GW observatory sensitive to gravitational signals between 0.1 mHz and 1 Hz, where we expect a large diversity of sources, ranging from supermassive black-hole binaries, quasi-monochromatic Galactic binaries, extreme mass-ratio inspirals, and stellar-mass binaries [3]. In addition to these expected sources, a number of potential signals might be detected, including stochastic GW signals from the early Universe, cusps and kinks of cosmic strings and other unmodeled burst sources. Precise measurements of the source parameters will help answer many astrophysical and cosmological questions, as well as constrain models beyond the general theory of relativity.

Gravitational waves perturb the curvature of spacetime, and LISA is able to measure them by monitoring the relative motion of inertial references at picometer precision, using heterodyne laser interferometry. The inertial references are realized by 6 cubic test masses, which are free-falling inside 3 spacecraft. These spacecraft will form a triangular constellation, which will orbit around the Sun. The main source of noise in downlinked LISA will be the phase (or frequency) instabilities of the laser sources, expected many orders of magnitude above the sensitivity target of the instrument. Contrary to ground-based interferometers, LISA has unequal (and time-varying) arm lengths, such that laser noise does not cancel when the laser beams recombine. To remedy this, time-delay interferometry (TDI) has been proposed in the early day of the project [4, 5]; by appropriately combining delayed copies of the interferometric measurements, it provides laser noise-free data channels. These are then limited by secondary, unsuppressed noises, mainly test-mass acceleration noise and interferometric readout noise, which ultimately set the sensitivity of the instrument. TDI has been studied extensively, and tested using both numerical simulations and table-top experiments [6–16].

TDI requires as input light travel times (LTTs), i.e. the values of the time-varying delays experienced by the laser beams when they propagate from one spacecraft to another[8]. To suppress laser noise to the required levels, each of these LTTs needs to be estimated with nanosecond precision, equivalent to knowing the spacecraft separation to around 30 cm [17]. Additionally, LTTs are needed to construct the response function of the instrument, which is used to estimate parameters of gravitational waves detected by LISA [18]. In this case, however, the required precision is much less stringent, at the level of microseconds (defined by the phase requirement on the reconstructed signal).

The baseline method to estimate these LTTs is to modulate the local laser phases with a megahertz-pseudorandom noise (PRN) code computed from the onboard clocks. The modulated beams are then received by the distant optical benches, and compared with a version of that signal generated according to the local clocks. This provides accurate estimates of the LTTs[9]. Note that the precision of these LTT estimates will be improved by fusing them with information contained in the clock-modulated sidebands [17].

The PRN system lacks any redundancy, and any failure, such as a loss of lock of the delay-lock loop, will result in the (potentially temporary) lack of LTT estimates. As a consequence, one will not be able to form a complete set of TDI combinations, translating directly into a loss in detection sensitivity [22]. In addition to this risk, the PRN-based estimated of the LTTs are intrinsically biased since the PRN sidebands experience optical and electronic delays that may be different from those experienced by the optical carriers; recent hardware experiments point to differences of hundreds to thousands of meters [23]. These carriers actually contain the GW signals and laser noise to be reduced by TDI. Calibration techniques to correct for part of the difference in these delays have recently been proposed [21].

However, an alternative method has been proposed: time-delay interferometric ranging (TDIR) provides bias-free estimates of the LTTs. TDIR can be used in case of PRN failure (as it requires no addition hardware); to estimate intrinsic biases of PRN-based estimates; or even as a sanity check for other LTT estimation methods [24, 25]. The idea behind TDIR is to estimate the LTTs by varying the delays in TDI combinations, and minimizing the residual laser noise over the entire LISA frequency band. More recently, it has been demonstrated that TDIR can be applied in a Bayesian framework, for a static or time-varying arm setup [26, 27].

Secondary noise (readout noise and test-mass acceleration noise) remains unsuppressed in the TDI combinations. They limit the laser noise minimization process, and therefore ultimately the precision of TDIR estimates. However, it is possible to obtain LTT estimates that reduce laser noise to below the level of secondary noises by accumulating more data. In short, the expected precision is driven by the laser noise-over-secondary noise level ratio (laser noise plays here the role of the signal) and averaging time.

Many GW signals will populate the LISA band. Some of these will be long-standing and relatively low signal-to-noise ratio (SNR) (such as the yearly-modulated confusion noise from millions of unresolved Galactic binaries, or potential cosmological stochastic gravitational background), while others will be very bright but remain in band only for weeks or months (such as supermassive back-hole binaries). In addition, data artifacts, such as spectral lines or glitches might affect the recorded measurements. Any feature in the data deviating from the noise models, including these gravitational signals and data artifacts, will degrade the performance of TDIR.

---

[8] In this paper, we ignore the relativistic drifts of the spacecraft proper times and imperfection of the clocks, which also enter the delays experienced by the laser beams, giving rise to the measured-pseudo ranges (MPRs) [17].

[9] PRN measurements actually provide estimates of the MPRs, but it has been shown that we can disentangle LTTs from clock drifts [20]. Because the PRN code has a finite length after which it repeats, the LTT are estimated up to the code length, expected to be around 400 km. This ambiguity can be resolved using ground-based observation of the spacecraft [21].

Following proposals made for a LISA and GRACE Follow-On TDIR demonstrations [14, 28], we propose in this paper to modulate the laser sources in LISA at one or a few specific frequencies in band, far above the expected frequencies of GW signals (0.1–1 Hz) to prevent interference. We dub this method modulation-assisted TDIR (MA-TDIR). MA-TDIR will increase the laser noise-over-secondary noise level ratio at those frequencies. The result is increased precision of the TDIR estimates (for a given averaging time) or a decreased averaging time necessary to reach required laser noise suppression performance. MA-TDIR also renders the method more robust to GW sources and other in-band artifacts, effectively decoupling the LTT estimation from the global fit.

The paper is structured as follows. In section 2, we use a toy model to introduce the concept of TDIR and MA-TDIR. We use the Fisher information matrix (FIM) formalism to compute the expected precision of the LTT estimates, and show how modulating the laser source can improve this precision. We move to the realistic laser-locked LISA case in section 3, discussing the potential degeneracies and solutions to break them. In section 4, we explore the feasibility of the various methods to inject tones in the laser sources and present the results of instrumental experiments that have been carried out. We then present in section 5 the results of our numerical simulations, and show that they match our analytical expectations. Lastly, we conclude in section 6.

## 2. TDIR on a toy model

In favor of a simplified setup, we first discuss TDIR and MA-TDIR for the toy model shown in figure 1. We then move on to the LISA case in the next section.

The toy model represents an unequal-arm Michelson interferometer with arm lengths $d_{121}$ and $d_{131}$, here expressed in seconds. The interferometric signals for both arms are read out individually by two photodetectors as differential phases,

$$\eta_{12}(t) = (\mathbf{D}_{121} - 1)\phi(t) + N_{12}(t), \tag{1a}$$

$$\eta_{13}(t) = (\mathbf{D}_{131} - 1)\phi(t) + N_{13}(t). \tag{1b}$$

Here, $\phi(t)$ denotes the total phase of the laser and $N_{ij}(t)$ readout noise of the respective photodetector. These terms also contain the GW signals; to make the equations more concise, we will not write out the signal contributions explicitly in most of this paper. Furthermore, we make use of the short-hand notation of the delay operation and introduce the delay operator defined as

$$\mathbf{D}x(t) = x(t-d). \tag{2}$$

The aim of TDIR is to extract the ranging information that is included in the interferometric measurements above.

We use the maximum likelihood approach and the FIM formalism to find an upper bound on the estimation variance of TDIR. Following the argument developed in [25], in the limit where laser noise power is much larger than secondary noise power, the entire information on the delay parameters is contained in a single data stream: the data combination that cancels laser noise

$$X(t) = \left(1 - \hat{\mathbf{D}}_{121}\right)\eta_{13}(t) - \left(1 - \hat{\mathbf{D}}_{131}\right)\eta_{12}(t). \tag{3}$$

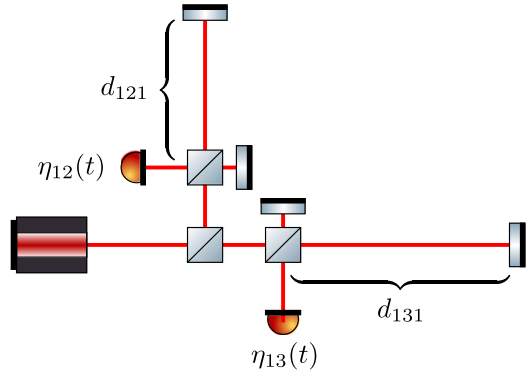

**Figure 1.** Illustration of the toy model used to simplify the LISA setup. Both arms of an unequal-arm Michelson interferometer are read out individually by two photodetectors. Note that we neglect the distance between the beam splitters.

Here, the hat on the delay operators indicates that we lack perfect knowledge of the delays and rely on estimates. In fact, the residual laser noise content in $X(t)$ is sensitive to the correct choice of delays. Conversely, the remaining laser noise-dominated combination is not sensitive to the choice of these delays. This provides an intuitive explanation why $X(t)$ holds all the information on the delays.

The logarithm of the stationary, zero-mean Gaussian likelihood is a function of parameters $\boldsymbol{\theta} = \begin{pmatrix} d_{121} & d_{131} \end{pmatrix}^{\mathrm{T}}$ given the data combination $X(t)$. It can be concisely written down (up to a constant) in Fourier space because frequency components become uncorrelated in the limit of long integration time $T$. We have

$$l(\boldsymbol{\theta}) \sim -\sum_{k=1}^{K} T \frac{|\tilde{X}_k|^2}{S_X(f_k)}. \tag{4}$$

Here, $\tilde{X}_k$ denotes the $k$th Fourier component of $X(t)$ at frequency $f_k = k \cdot \Delta f = k/T$ and $S_X(f)$ the power spectral density (PSD) of secondary noises in $X(t)$. The latter is given by

$$S_X(f) = 4 \left( \sin^2 \left( \pi f d_{121} \right) + \sin^2 \left( \pi f d_{131} \right) \right) \cdot S_N(f), \tag{5}$$

where we assume that the readout noise on the photodetectors, as seen in equation (1), is independent and identically distributed, with a PSD of $S_N(f)$. In general, some of the LISA noise sources are expected to be potentially slightly correlated across the interferometric measurements. This can be accounted for by cross-terms in the model of the PSD $S_N(f)$. This will be the ultimate noise floor that limits the precision of the estimate $\boldsymbol{\theta}$.

From the log-likelihood function given in equation (4) we can directly write down the FIM. Its entries are defined as

$$\mathrm{I}_{i,j} = -\mathrm{E} \left\{ \frac{\partial^2 l(\boldsymbol{\theta})}{\partial \theta_i \partial \theta_j} \right\}. \tag{6}$$

For the calculation we assume that the derivative of the total laser phase, i.e. the total laser frequency $\nu(t)$, is equal to the central laser frequency $\nu_0$ and additive laser frequency noise $\dot{p}(t)$. The PSD of laser frequency noise $S_{\dot{p}}(f)$ before TDI is much larger than the PSD of secondary noises $S_N(f)$ over the entire band. Furthermore, we simplify the expression for long integration times $T$ to turn the discrete sum over frequency bins into an integral and equal arms. We find

$$I = T \int \frac{S_{\dot{p}}(f)}{S_N(f)} \begin{pmatrix} 1 & -1 \\ -1 & 1 \end{pmatrix} df, \tag{7}$$

which is singular (has determinant of zero). This follows from the fact that in the equal-arm case TDIR cannot measure one particular linear combination of delays; the sum $d_{131} + d_{121}$. Instead, it can only provide an estimate for the differential delay $\Delta d = d_{121} - d_{131}$. We can determine the Cramér–Rao lower bound on its estimation variance by inverting the Fisher information and find

$$\sigma_{\Delta d} \geqslant \sqrt{\frac{1}{T} \left( \int \frac{S_{\dot{p}}(f)}{S_N(f)} df \right)^{-1}}$$
$$\sim 120 \, \text{m} \cdot \left( \frac{T}{0.25 \, \text{s}} \right)^{-\frac{1}{2}}. \tag{8}$$

In the second line we evaluate the expression for realistic parameters by computing the integral with an upper limit of 1 Hz. We assume constant amplitude spectral densities (ASDs) for laser frequency noise and secondary noises given by $\sqrt{S_{\dot{p}}} = 30 \, \text{Hz} \sqrt{\text{Hz}^{-1}}$ and $\sqrt{S_N} = 6 \, \mu\text{cycles} \sqrt{\text{Hz}^{-1}}$. For easy comparison with the PRN accuracy we choose a characteristic integration time of 0.25 s, which coincides with the sampling rate of the PRN measurement, and express the prefix in units of light-meters.

In this paper we suggest to modulate a tone on the lasers. Artificially increasing the laser noise power in a narrow band enables us to decrease the integration time $T$ and to become more robust against estimation biases from GW signals and glitches. In the case of a single strong tone, the sum in the log-likelihood in equation (4) reduces to a single component, the one at the tone's frequency $f_t$. As a result the estimation variance simplifies to

$$\sigma_{\Delta d} \geqslant \sqrt{\frac{1}{T} \frac{S_N(f_t)}{P_t}} \sim 5 \, \text{m} \cdot \left( \frac{T}{0.25 \, \text{s}} \right)^{-\frac{1}{2}}, \tag{9}$$

where the tone power $P_t = \frac{A_t^2}{2}$ is dependent on the tone amplitude $A_t$. For the numerical evaluation of the expression we use a tone amplitude of 1 kHz and the same level of secondary noises as before. Note that, in this case, the variance is independent of the tone frequency $f_t$. However, we suggest to place the tone at high frequency, close to the LISA band edge, to avoid spectral overlap with GW sources. We observe that the introduction of a kilohertz-tone allows to reduce the standard deviation of the estimator by a factor of 24. However, this remains less accurate than the PRN-based estimate (1 m RMS).

## 3. TDIR with LISA

Let us now turn to the full LISA setup, which increases the complexity of the problem but can be considered an extension of the toy model. The LISA constellation is composed of three spacecraft, each of which hosts two movable optical sub-assemblies (MOSAs). Each MOSA holds a laser and an optical bench that receives light from the distant and the adjacent MOSA.

Beatnote phases are recorded in the inter-spacecraft interferometers and the reference interferometers[10],

$$\text{isi}_{ij} = \mathbf{D}_{ij}\phi_{ji} - \phi_{ij} + N_{ij}, \tag{10}$$
$$\text{rfi}_{ij} = \phi_{ik} - \phi_{ij}. \tag{11}$$

Here, $ij$ denotes the index pair of the local MOSA, where $i$ is the index of the hosting spacecraft and $j$ the index of the distant spacecraft. We only consider secondary noise in the inter-spacecraft interferometer as it is the dominant contribution to the LISA noise budget. Consequently, adjacent reference interferometers become redundant and we can also limit ourselves to those on left-handed MOSAs (i.e. with indices 12, 23, and 31). This leaves us with a total of nine phasemeter equations involving six lasers.

The six lasers in LISA are stabilized via two methods. Each MOSA hosts an optical cavity that is used as a frequency reference to stabilize the laser's central frequency to the usual $30\,\text{Hz}\,\sqrt{\text{Hz}^{-1}}$ in-band level. However, the LISA heterodyne phase readout system requires the relative laser frequencies to fall into a specific range of 5–25 MHz. To ensure this only a single laser, in the following called the 'primary laser', is locked to its cavity. The remaining lasers are frequency-offset locked to the primary laser according to a frequency plan. Frequency-offset locks are established by acting on the frequency $\nu_{ij} = \dot{\phi}_{ij}$ of the locked laser so that the beatnote frequency measured in the respective interferometer follows a predetermined offset $o_{ij}$. When a laser is locked to the distant (resp. adjacent) laser, the inter-spacecraft interferometer (resp. reference interferometer) is used. This procedure leads to the so called locking conditions [30, 31],

$$\dot{\text{isi}}_{ij} = o_{ij} \quad\Rightarrow\quad \nu_{ij} = \mathbf{D}_{ij}\nu_{ji} + N_{ij} - o_{ij}, \tag{12}$$
$$\dot{\text{rfi}}_{ij} = o_{ij} \quad\Rightarrow\quad \nu_{ij} = \nu_{ik} - o_{ij}. \tag{13}$$

We observe that, due to the frequency-offset locks, laser frequency noises measured in all interferometers are correlated.

Constructing laser noise-free combinations for the full LISA case involves an additional step. As LISA hosts a total of six lasers, we remove three of them (lasers on right-handed MOSAs, with indices 13, 32, and 21) by forming the intermediary variables $\eta_{ij}$ [32]. They are defined as

$$\eta_{ij} = \begin{cases} \text{isi}_{ij} - \hat{\mathbf{D}}_{ij}\text{rfi}_{jk} & \text{if } \epsilon_{ijk} = 1, \\ \text{isi}_{ij} + \text{rfi}_{ik} & \text{if } \epsilon_{ijk} = -1, \end{cases} \tag{14}$$

where $\epsilon_{ijk}$ is the Levi–Civita symbol, which is 1 (resp. $-1$) for left-handed (resp. right-handed) MOSAs. As mentioned before, we only make use of the reference interferometers on left-handed MOSAs in the previous equations and adjust the definition of $\eta_{ij}$ accordingly.

Next, we construct three TDI variables $X$, $Y$ and $Z$, which span the full laser noise free space in LISA. They are given by

$$X = \left(1 - \hat{\mathbf{D}}_{121}\right)\left(\eta_{13} + \hat{\mathbf{D}}_{13}\eta_{31}\right) - \left(1 - \hat{\mathbf{D}}_{131}\right)\left(\eta_{12} + \hat{\mathbf{D}}_{12}\eta_{21}\right), \tag{15}$$

---

[10] We omit the definition of the test-mass interferometer as it is redundant if optical bench motion and test-mass acceleration noise are neglected.

where $Y$ and $Z$ can be derived by cyclic permutation of the indices ($1 \to 2$, $2 \to 3$ and $3 \to 1$). We give the expressions for the TDI generation 1.5, which exactly cancel laser frequency noise for a static constellation with six unequal arms.

Analogous to the toy model presented in section 2, we derive the FIM for various injection scenarios. To account for the correlations between the three laser-noise-free combinations $X$, $Y$, and $Z$, the likelihood function $l(\boldsymbol{\theta})$ turns into a multivariate Gaussian likelihood. This likelihood is a function of the random vector of Fourier components $\tilde{\boldsymbol{X}}_k = \begin{pmatrix} \tilde{X}_k & \tilde{Y}_k & \tilde{Z}_k \end{pmatrix}^\mathrm{T}$ at frequency $f_k$, with covariance matrix

$$\Sigma_X(f_k) = \frac{1}{T} \begin{pmatrix} S_{XX}(f_k) & S_{XY}(f_k) & S_{XZ}(f_k) \\ S_{YX}(f_k) & S_{YY}(f_k) & S_{YZ}(f_k) \\ S_{ZX}(f_k) & S_{ZY}(f_k) & S_{ZZ}(f_k) \end{pmatrix}. \tag{16}$$

Here, the diagonal and off-diagonal entries represent power spectral and cross spectral densities, respectively. As $\Sigma_X$ only models the secondary noise content, the terms $S_{XX}(f)$ and $S_{XY}(f)$ have the usual symmetry properties under cyclic permutation of the indices if equal readout noise levels in all inter-spacecraft interferometers are assumed.

The general FIM, however, does not share this symmetry property. Due to laser locking the three laser-noise-free combinations have rather dissimilar responses to errors in the delay estimates (see [9]), which leads to different estimation variances of the six delays. A general expression for the Cramér–Rao bound for standard TDIR (full-band) is given by

$$\Sigma_d \geqslant \frac{1}{T} \left( \int \frac{S_{\dot{p}}(f)}{S_N(f)} \Lambda(f) \, \mathrm{d}f \right)^{-1}, \tag{17}$$

where the matrix $\Lambda(f)$ is dependent on the locking configuration used and the value of the six delays. As opposed to the toy model, the integral evaluates to an invertible matrix for equal arms. Therefore, the previous degeneracy is lifted by including the combinations $Y$ and $Z$. In this case, all six arms can be determined individually.

To derive further properties and numerical estimates of the Cramér–Rao bound we have to make assumptions on the setup. First, we choose a particular locking configuration; we pick the mission baseline configuration N1-L12 (topology N1 with primary laser 12) [30]. Secondly, we pick a set of constant delays[11] from a realistic numerical orbit file provided by ESA [33], at some representative arbitrary time. They read

$$d_{12} = 8.171\,\mathrm{s}, \qquad d_{23} = 8.323\,\mathrm{s}, \qquad d_{31} = 8.243\,\mathrm{s},$$
$$d_{13} = 8.243\,\mathrm{s}, \qquad d_{32} = 8.322\,\mathrm{s}, \qquad d_{21} = 8.173\,\mathrm{s}.$$

Furthermore, we assume identical levels as in the toy model for laser frequency noise of the primary laser and readout noise in the inter-spacecraft interferometers. We set the integration time to 0.25 s for easy comparison with the PRN performance.

In figure 2 we present the performance of TDIR for different scenarios. Each plot shows a representation of the inverse of the FIM $\Sigma$; the Cramér–Rao bound. The main diagonal reads the standard deviation $\sigma_i = \sqrt{\Sigma_{i,i}}$ of the individual delay estimates in units of light-meters. They can directly be compared with the RMS performance of the PRN measurement and the pre-factors in equations (8) and (9). The color-coded off-diagonal terms show correlation coefficients, which are computed as $\Sigma_{i,j}/\sigma_i \sigma_j$.

---

[11] Note that, in reality, these delays are not constant, and the spacecraft relative velocities induce Doppler shifts in the tones we inject. A quick calculation shows that this shift corresponds to $\Delta\nu \approx 1\,\mathrm{Hz} \times 10\,\mathrm{m\,s^{-1}}/c \approx 1\,\mathrm{yr}^{-1}$, and therefore can be neglected.

**Table 1.** Arm sensitivities when injecting a single tone in various lasers. Results are shown here for the mission baseline locking configuration N1-L12. Tick marks (✓) indicate that a frequency modulation in the respective laser enables us to measure a certain delay while crosses (✗) indicate complete insensitivity.

| N1-L12 | $d_{12}$ | $d_{23}$ | $d_{31}$ | $d_{13}$ | $d_{32}$ | $d_{21}$ |
|---|---|---|---|---|---|---|
| $L_{12}$ | ✓ | ✓ | ✓ | ✓ | ✓ | ✓ |
| $L_{23}$ | ✗ | ✗ | ✗ | ✗ | ✓ | ✗ |
| $L_{31}$ | ✗ | ✓ | ✗ | ✓ | ✗ | ✗ |
| $L_{13}$ | ✗ | ✓ | ✓ | ✓ | ✗ | ✗ |
| $L_{32}$ | ✗ | ✓ | ✗ | ✗ | ✗ | ✗ |
| $L_{21}$ | ✓ | ✗ | ✗ | ✗ | ✓ | ✗ |

The upper left panel of figure 2 shows the result for standard TDIR. As already stated before, individual delays can be determined due to the inclusion of *Y* and *Z*. Note that delay estimates are correlated. In particular, reciprocal delay estimates $d_{12}$ and $d_{21}$ (resp. $d_{13}$ and $d_{31}$) are anti-correlated; this is due to the chosen locking configuration, which exhibits a symmetry around spacecraft 1.

To improve the performance of the TDIR estimates we propose to add modulations on the laser frequencies. In table 1 we list the arm sensitivities for six possible injection points in LISA; the primary laser or the five locking beatnotes (the latter being easier to do from an instrumental point of view, see section 4). We see that modulating the primary laser is the only option to provide estimates for all six delays. For modulation of the five remaining lasers the likelihood function is insensitive to at least one of the delays. Here, delay $d_{21}$ stands out as it can only be recovered modulating the primary laser. Therefore, we present in the upper right-hand correlation matrix of figure 2 as the most simple case the injection of a single 1 kHz tone in the primary laser, at a frequency of $f_0 = 0.94\,\text{Hz}$. The estimates have standard deviations of approximately 10 light-meters and are therefore not competitive with PRN estimates. Furthermore, they show strong correlations, which turn into full degeneracy along a specific 'direction' when arms become equal. The insensitive direction is defined by the linear combination of delays

$$(1+2a)(d_{12}+d_{13}) + 2a(d_{23}+d_{32}) + (d_{31}+d_{21}), \tag{18}$$

where $a = \cos(2\pi f d)$ is a frequency-dependent factor.

To further improve the precision and reduce the correlations of the delay estimates, we study the injection of two tones in the primary laser at frequencies $f_0 \pm \Delta f/2$. We set $f_0 = \frac{2n+1}{4d}$ to fall on a TDI transfer function maximum (where *n* is an integer and denotes the *n*th maximum) and optimize $\Delta f$ such that the corresponding degenerate directions are nearly orthogonal. We find that $\Delta f = \frac{1}{6d} \approx 0.02\,\text{Hz}$ for an average arm length of $d = 8.3\,\text{s}$. We observe that the standard deviations of the delay estimates is below 10 light-meters and correlations are reduced (see lower left-hand panel of figure 2).

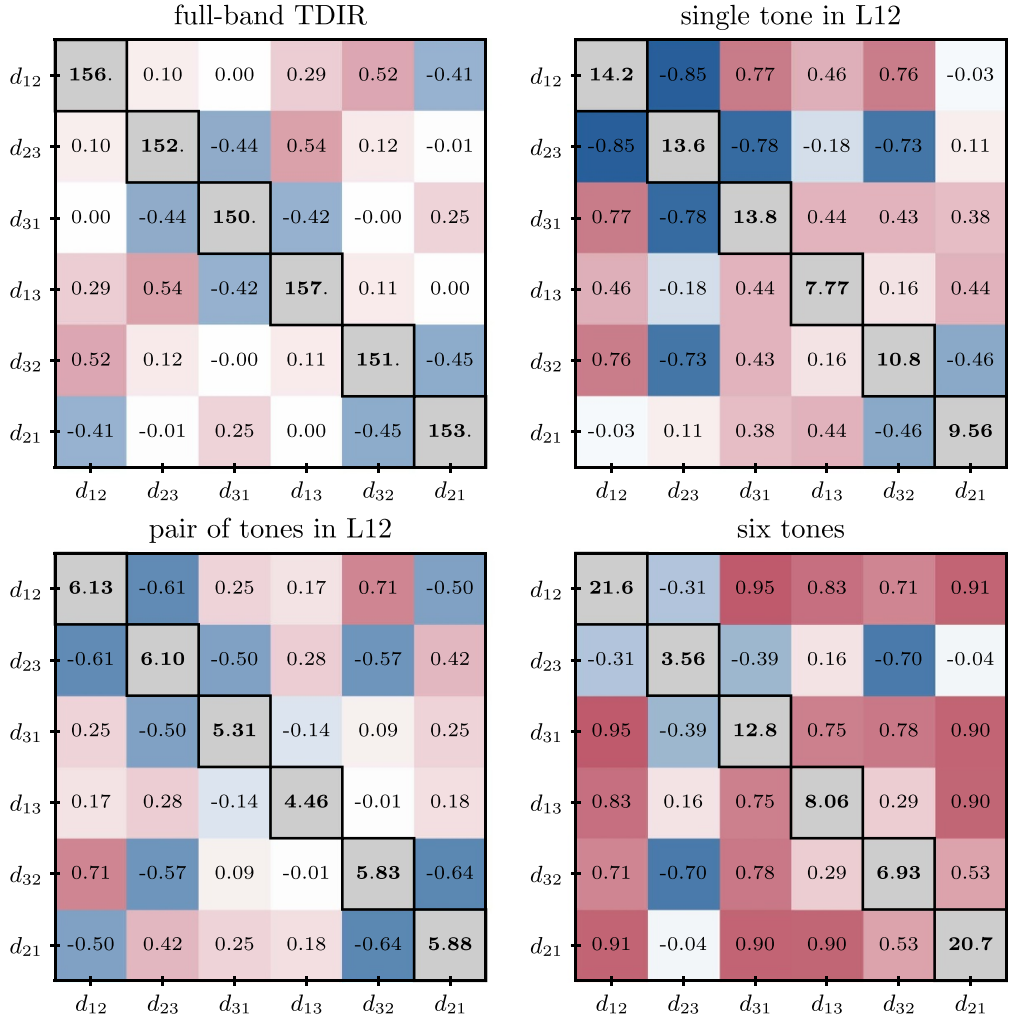

**Figure 2.** Correlation matrices for standard full-band TDIR (relying on the natural laser frequency noise), when injecting a single tone at $f_0$, when injecting two tones at $f_0 + \frac{\Delta f}{2}$ and $f_0 - \frac{\Delta f}{2}$, and when injecting the same tone at $f_0$ in all six locks. Diagonal elements are the standard deviation of the respective estimate for an integration time $T = 0.25\,\text{s}$. Results are shown here for the mission baseline locking configuration N1-L12.

Lastly, we compute the correlation matrix for a single tone injected in all six lasers, at the same frequency $f_0$, and show the results in the last plot of figure 2. We observe that the precision of the estimates is not improved compared to the two-tone case. This is due to non-trivial interactions between the modulations from the six lasers, which can lead to destructive interference on single-link $\eta$ measurements, dependent on the chosen locking scheme, tone initial phases and frequency $f_0$, and the values of the delays. Furthermore, correlations between the delay estimates are strongly increased and are much greater that those obtained with standard TDIR.

Note that, because we have non-vanishing correlations in our correlation matrices, we cannot formally compare the errors on the delay estimates obtained through MA-TDIR and those obtained using the baseline PRN method (which we can take, at first order, as uncorrelated). However, we have computed the associated eigenvalues in the 4 scenarios (full band, single tone, two tones, and six tones), and found the same orders of magnitudes as the diagonal elements,

$$\sqrt{\lambda_{\text{full band}}} \approx \{229.06, 195.86, 143.33, 114.04, 94.81, 86.23\},$$
$$\sqrt{\lambda_{1\text{ tone}}} \approx \{23.92, 13.02, 7.65, 4.16, 3.8, 3.68\},$$
$$\sqrt{\lambda_{2\text{ tones}}} \approx \{9.98, 6.45, 5.18, 3.07, 2.71, 2.56\},$$
$$\sqrt{\lambda_{6\text{ tones}}} \approx \{32.8, 8.63, 4.7, 2.4, 2.03, 1.57\},$$

As a consequence, all previous discussions remain valid.

As a natural conclusion, we suggest to inject two tones at well chosen frequencies for a robust TDIR implementation.

We investigate the five remaining locking configurations [30]. We find that, under the equal-arm assumption and for a single-tone injection, N2 and N4 have a similar degeneracy as N1. In a realistic unequal-arm setup, this degeneracy translates into correlated delay estimates. Conversely, N3, N5, and N6 yield full-rank FIMs, indicating that we can determine all 6 delays in the equal-arm case.

As a cross-check, we compare our results with the existing TDIR literature, which only considers the simplified case of six unlocked lasers. Using the associated diagonal FIM computed in [25], we find that full-band TDIR yields delay estimates with a precision of $\sim$120 m (using the same integration time of 0.25 s) [24] reports a precision of $\sim$400 m; the order of magnitude is similar, and differences can be explained by the different assumptions on noise levels. Numerical experiments in [26] yield a precision of $\sim$1900 m. However, they integrate over a small band, corresponding roughly to a tenth of the usual LISA band; scaling the number accordingly, we find the correct order of magnitude.

## 4. Instrumental considerations

In LISA, the primary laser will be frequency-stabilized by locking its frequency to an optical cavity using the Pound–Drever–Hall (PDH) locking method. It is possible to add a frequency modulation to the primary laser by adding a modulation signal into the PDH control loop, at the error point of the laser locking controller, as shown in figure 3. A 100 mHz to 1 Hz laser frequency modulation is well within the bandwidth of typical PDH and offset-locking control loops (10 kHz) and the frequency response of the laser (100 kHz) [34]. Injecting MA-TDIR tones into the PDH control loop therefore would not require any modifications to the laser frequency stabilization hardware, only requiring a modification to firmware and flight software.

As discussed in the previous sections, MA-TDIR performs better with larger modulation depths. However there are both instrument and constellation constraints that bound the modulation parameters. The modulation waveform, frequency and amplitude are constrained by three main requirements: (1) the modulation must not break the PDH loop; (2) the modulation must not degrade laser frequency noise stability above the required $30\,\text{Hz}\,\sqrt{\text{Hz}^{-1}} \times \text{NSF}(f)$ level; and (3) the modulation frequency must be above the science signal band ($> 100\,\text{mHz}$), within the LISA band ($\leqslant 1\,\text{Hz}$), and not in a TDI null.

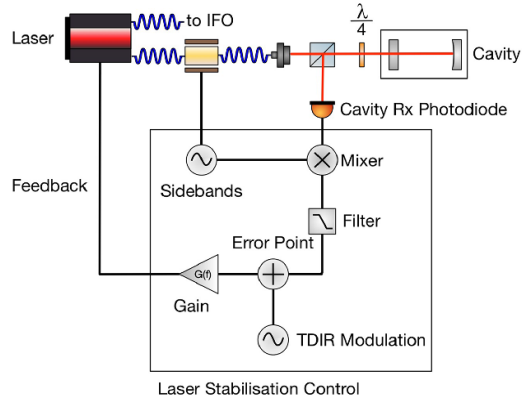

**Figure 3.** Proposed method to modulate primary laser with tone for MA-TDIR. The laser going to the interferometer will be frequency-stabilized and modulated.

The added modulation will not break the PDH loop if the amplitude is sufficiently small, such that the frequency deviations are within the linear region of the PDH error signal. The LISA cavity is based off the GRACE Follow-On optical cavity design, which had a linewidth of 150 kHz [35]. A 1 kHz-amplitude tone meets this criteria.

Similarly, keeping the tone amplitude small minimizes impacts to the laser frequency noise stability. A PDH cavity lock is first-order insensitive to optical power fluctuations [36]. Adding a tone to the laser pulls the laser frequency off cavity resonance, introducing sensitivity to relative intensity noise.

To verify that a 1 kHz tone at 100 mHz does not impact the performance of a PDH laser stabilization loop, an optical benchtop test was performed. This test, with results shown in figure 4, used a 12.5 k-finesse cavity, with the same geometry as the GRACE Follow-On and LISA cavities. In this test, a 1064 nm Lightwave NPRO was PDH-locked to the cavity using a commercial laser lock box. A 100 mHz tone, of varying amplitude, was injected into the PDH loop at the error point. To establish the laser frequency noise in the cavity locked laser, the laser was interfered with a second laser, locked to a different cavity. The frequency noise in the resulting beatnote was tracked with a phasemeter, resulting in the measurement spectra shown in the figure. As the second cavity was higher finesse, the error in the beatnote can be attributed to the laser locked to the LISA-like cavity. With a 10 kHz amplitude tone, the laser frequency stability is degraded and harmonics of the tone frequency appear due to nonlinearities as the laser frequency is pulled off cavity resonance. Different modulation frequencies were not explicitly tested but are expected to have similar results given the frequency response of lasers and cavity locking loops are flat within the LISA measurement band with bandwidths of order 10 kHz [34].

In addition to injecting a tone into the primary laser, other modulation schemes could be implemented. Injecting two tones into the primary laser has already been discussed as a solution to degeneracies in TDIR when arm lengths are close to equal. Another method that has been shown to break the degeneracies is, in addition to the tone in the primary laser, injecting a tone of the same frequency and amplitude into one or more of the phase-locked loops used to stabilize the frequency noise in the other lasers in the constellation. Injecting a tone in one or more phase-locked loop has the advantage of breaking degeneracy without adding an additional tone in the measurement band. Given the phase-locked loops used in phasemeters to derive their error signals, which are highly linear and have large dynamic range, injecting a tone in a phase-locked loop does not have the amplitude constraints of the tones injected into PDH.

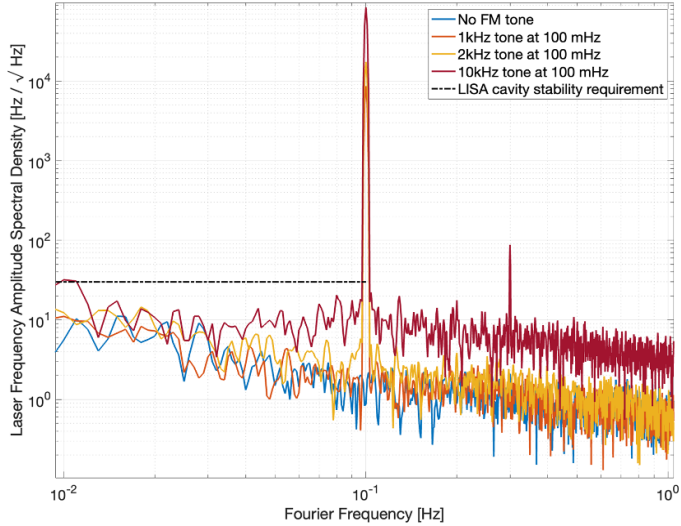

**Figure 4.** A benchtop cavity locking experiment was setup to verify injecting a 1 kHz amplitude tone at 100 mHz did not limit the performance of PDH. The amplitude of the tone was increased to show larger tones do degrade performance, increasing the overall noise around the tone.

Although we propose to place any tones above the GW signal band and have experimentally demonstrated that the tones would not degrade the performance of the laser frequency stabilization, the MA-TDIR concept could be achieved using a modulation that does not appear in the spectrum of the science measurement. This would use a spread-spectrum modulation, applied to the primary laser at the PDH error point. Similar to the PRN baseline that provides the LTTs, but with a significantly lower chip rate (100 mHz instead of 1 MHz), a spread spectrum modulation has the advantage of appearing below the shot noise level in the phasemeter measurements but still provides a strong calibration signal for the TDIR algorithm as PRN codes provide strong signals when auto-correlated over many chips.

## 5. Numerical simulations

To validate the analytical results described in section 3, we run numerical experiments. We use LISA Instrument [31] to simulate interferometric data (namely the inter-spacecraft, test-mass, and reference measurements), and PyTDI [37] to form the TDI variables given a set of delays.

We use most of the instrumental assumptions made in section 3. Interferometric beatnotes are formed using the baseline locking configuration N1 with primary laser 12, with a representative set of constant delays,

$$d_{12} = 8.171\,\text{s}, \qquad d_{23} = 8.323\,\text{s}, \qquad d_{31} = 8.243\,\text{s},$$
$$d_{13} = 8.243\,\text{s}, \qquad d_{32} = 8.322\,\text{s}, \qquad d_{21} = 8.173\,\text{s}.$$

We include laser, test-mass, and readout noise[12] in our simulations. Their respective PSDs are given by [31]

$$S_{\dot{p}}(f) = \left(30\,\text{Hz}\,\sqrt{\text{Hz}^{-1}}\right)^2,$$

$$S_{\text{TM}}(f) = \left(2.4\,\text{fm}\,\text{s}^{-2}\,\sqrt{\text{Hz}^{-1}}\right)^2 \left[1 + \left(\frac{0.4\,\text{mHz}}{f}\right)^2\right],$$

$$S_{\text{OMS}}^{\text{ISI}}(f) = \left(6.35\,\text{pm}\,\sqrt{\text{Hz}^{-1}}\right)^2 \left[1 + \left(\frac{2\,\text{mHz}}{f}\right)^4\right].$$

One can compute how these noises appear in the interferometric measurements by multiplying these expressions by the relevant transfer functions, e.g. as provided in [38]. As proposed in section 3, one or two tones of 1 kHz are injected in the lasers, at $f_0 = 0.94\,\text{Hz}$ or $f_0 \pm \Delta f/2$, with $\Delta f \approx 0.02\,\text{Hz}$. We simulate one day of data at 4 Hz, which is about $3.5 \times 10^5$ samples (note that we trim some samples at the beginning and end after computing TDI combinations to remove fractional delay filter edge effects).

We use the `scipy.optimize.minimize` optimizer to find the maximum-likelihood solution $\hat{\boldsymbol{\theta}}$. We compute the stationary multivariate Gaussian likelihood as

$$l(\boldsymbol{\theta}) \sim -\tilde{\boldsymbol{X}}(\boldsymbol{\theta},f_0)^{\text{T}} \Sigma_X^{-1}(f_0) \tilde{\boldsymbol{X}}(\boldsymbol{\theta},f_0), \tag{19}$$

or as the sum of two similar terms, replacing $f_0$ by $f_0 \pm \Delta f/2$. Here,

$$\tilde{\boldsymbol{X}}(\boldsymbol{\theta},f_0) = \begin{pmatrix} \tilde{X}_1(\boldsymbol{\theta},f_0) & \tilde{Y}_1(\boldsymbol{\theta},f_0) & \tilde{Z}_1(\boldsymbol{\theta},f_0) \end{pmatrix}^{\text{T}} \tag{20}$$

is the vector of Fourier transform estimates for first-generation TDI $X_1, Y_1$, and $Z_1$ combinations at $f_0$, computed with the proposed set of delays $\boldsymbol{\theta}$. Note that we use a taper window function $w(t)$ to compute these frequency-domain estimates, e.g.

$$\tilde{X}_1(\boldsymbol{\theta},f_0) = \int_0^{1\,\text{day}} X_1(\boldsymbol{\theta},t)\, w(t)\, e^{-i2\pi f_0 t}\, \text{d}t. \tag{21}$$

The secondary noise covariance matrix $\Sigma_X(f_0)$ is also computed at $f_0$ and for an initial guess of the delays. Because it only weakly depends on the delays $\boldsymbol{\theta}$, we neglect this dependency in our likelihood [25, 27].

To check that our delay optimizer correctly converges when using two-tone MA-TDIR, we show in figure 5 the ASDs of Michelson TDI combination $X_1$, calculated for three different sets of delays. The blue line is computed using the true delays; laser noise suppression is optimal in this case, and we are left with limiting, secondary (unsuppressed) readout and test-mass noise. In red, we show $X_1$ computed with initial parameters, which were randomly drawn around the true values from a normal distribution with a standard deviation of 1 $\mu$s. We clearly see that laser noise is not well suppressed over most of the band; as expected, the two tones are visible around $f_0 = 0.94\,\text{Hz}$. Lastly, the yellow trace (behind the blue trace in most of the frequency band) is computed using the set of estimated delays (maximum likelihood for 2 injected tones). We see that the tones are suppressed *at least* to the level of secondary noises (blue line), as is laser noise over the whole band. We deduce that our optimizer behaves as expected.

---

[12] We only consider readout noise in the inter-spacecraft interferometers, as it is the dominant contribution to the LISA noise budget.

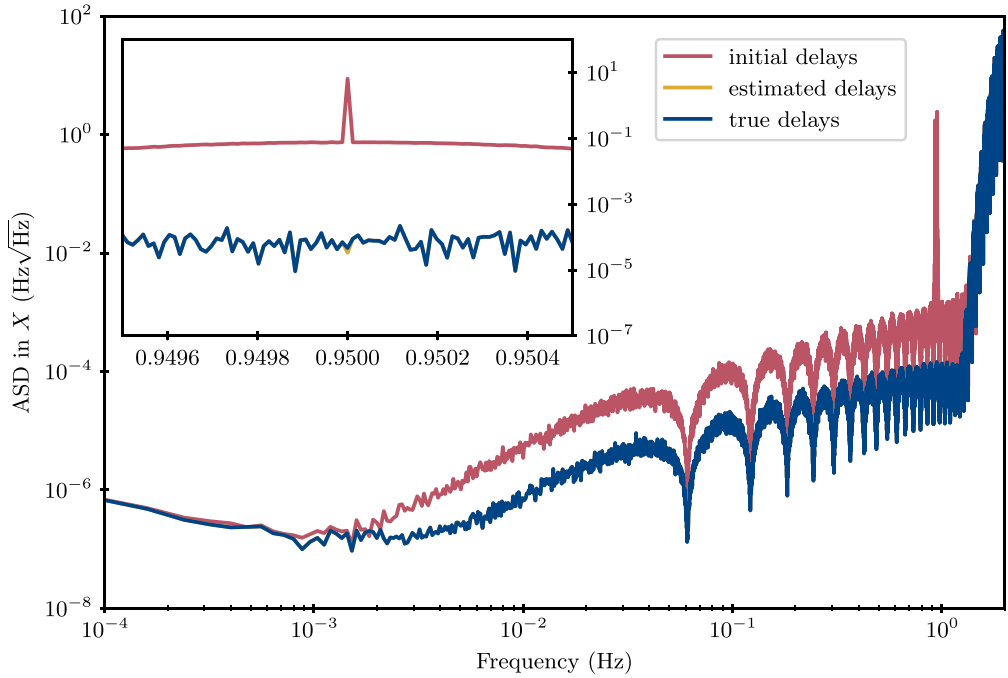

**Figure 5.** ASDs of $X_2$ computed with initially proposed parameters (see the two tones clearly above the residual noises), with optimized delays (no tones visible), and one with true delays (no tones either). The upper left panel shows a detailed view around one of the tone frequencies.

To estimate the properties of our estimation method, we run the optimizer on an ensemble of 10 000 datasets[13], each with independent noise realizations. For each dataset, we use the same likelihood and optimizer parameters. To speed up convergence, we use the true delays as the initial guess. Figure 6 shows the distribution of the maximum-likelihood estimates when injecting one tone (in red) or two tones (in blue). The numbers are rescaled to be easily compared with the theoretical correlation matrices in figure 2. We find that these empirical results are in very good agreement with the theoretical expectations: correlations are significantly reduced when using a pair of modulating tones and, as a consequence, the uncertainty in this case on the delay estimates are about 6 m for an integration time of 0.25 s.

In the introduction, we claim that MA-TDIR is more robust against in-band signals or data artifacts than the traditional TDIR approach. We now demonstrate this robustness using numerical simulations under a simple scenario. We run simulations on an ensemble of 10 000 datasets, first using the traditional TDIR approach; then with two-tone MA-TDIR injections at $f_0 \pm \Delta f/2 \approx 0.94$ Hz. In each case, we compare the maximum-likelihood delay estimate distributions with only laser and secondary noises (our baseline) and when one short, loud glitch fitted from LISA Pathfinder data [39] is injected on top of these noises[14]. The glitch spectrum is mostly white at low frequencies, clearly peaking above the secondary noises, as shown in figure 7.

---

[13] This choice ensures significant statistics as it yields, per the central limit theorem, relative standard deviations of the estimation variance and bias of $\sqrt{2/(N-1)} \approx 1.4\%$ and $1/\sqrt{N} = 1\%$, respectively.

[14] For this proof-of-principle demonstration, we amplified the amplitude of the glitch by a factor of 10 to emphasize its impact.

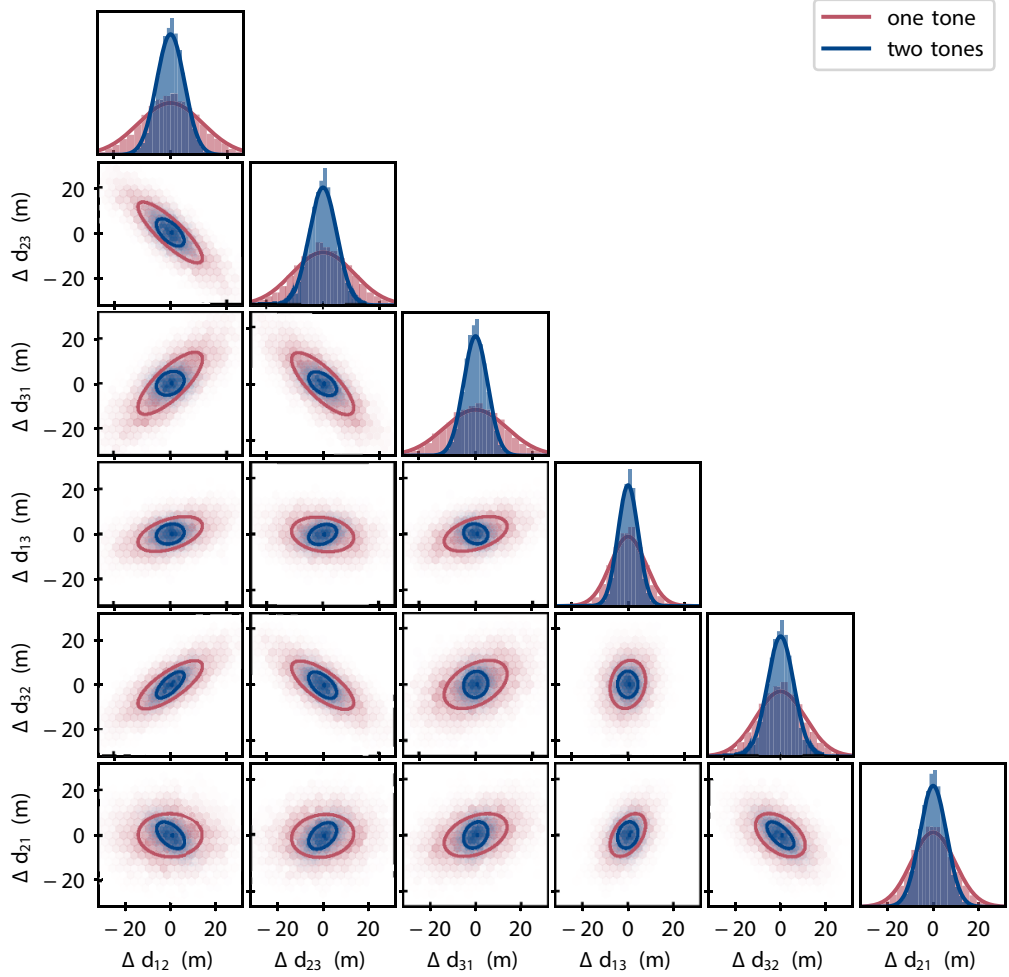

**Figure 6.** Distributions of maximum-likelihood delay estimates for an ensemble of 10 000 datasets, injecting one tone at $f_0$ (in red) and a pair of tones at $f_0 \pm \Delta f/2$ (in blue). Dark red and blue solid lines show theoretical marginal distributions and $1-\sigma$ confidence interval regions for the one and two tone cases, respectively.

The results of these simulations are presented as violin plots in figure 8. In the upper panel, we present the 6 delay estimate distributions obtained using the full-band likelihood (computed in the frequency domain, summing equation (19) over all frequencies between $10^{-4}$ Hz and 1 Hz) [26]. The lower panel shows these distributions when using the two-tone likelihood. On the left-hand side, blue distributions are obtained with laser and secondary noises only, while red distributions show the distributions when we inject the glitch in the data. We clearly see that, in the full-band case, the estimate variance is strongly degraded for all links when the glitch is present in the data; on the other hand, in addition to a reduced variance (expected from the theoretical development presented in section 3), the MA-TDIR estimate spread remains unchanged in the presence of the data artifact.

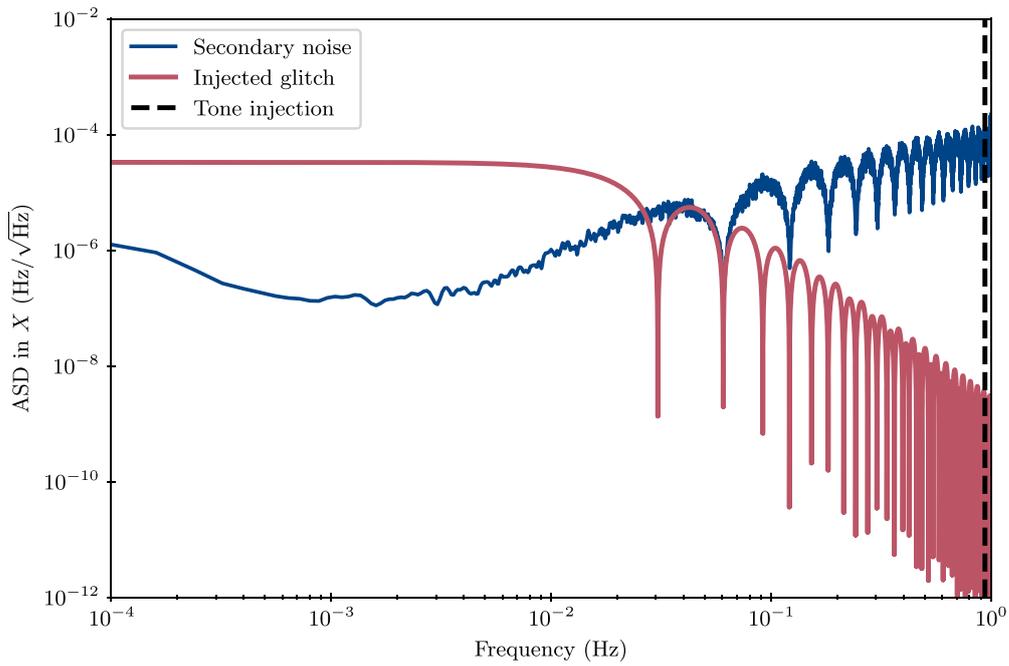

**Figure 7.** Injected glitch periodogram (in red) and secondary noise ASD (in blue). The vertical dashed line indicate the tone injection frequency $f_0$.

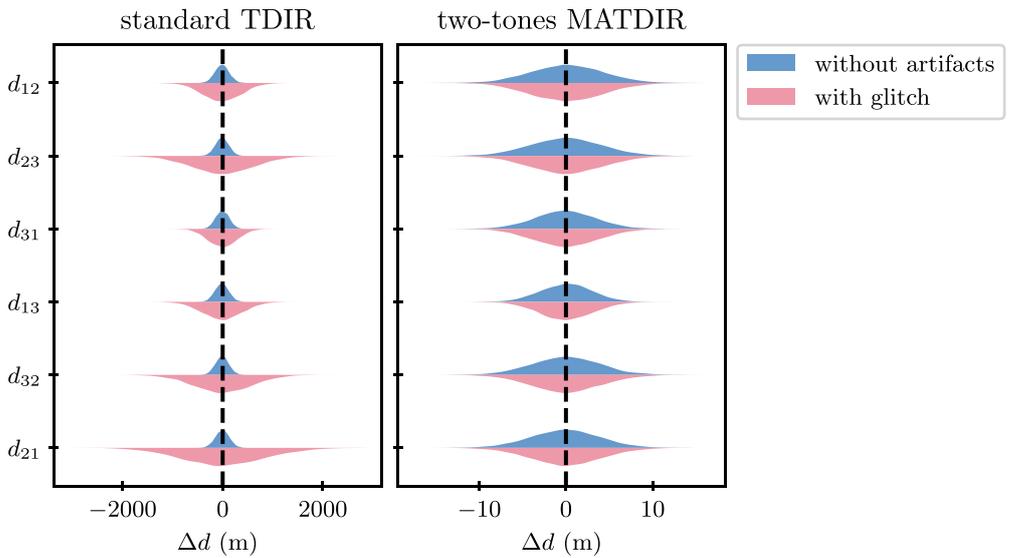

**Figure 8.** Marginal distributions of the six arms without artifacts (in blue) and with glitch (red). The upper plot shows the results for standard TDIR and the lower of MA-TDIR. Note the different *y*-axis scale, as MA-TDIR delay uncertainties are much reduced compared to those obtained using standard TDIR.

The glitch signal acts as an additional, non-suppressed power below $\approx 2 \times 10^{-2}$ Hz, and, similarly to secondary noises ultimately limits the precision of TDIR estimates. Because MA-TDIR only accounts for power around the injected tone frequency $\approx 1$ Hz they are mostly insensitive to this artifact. Note that, even if unsuppressed power is present around the MA-TDIR tone frequencies, the higher 'laser-noise-to-unsuppressed-power ratio' ensures that we obtain more precise results.

## 6. Conclusion and outlook

Estimating the photon time of flight between the spacecraft of a space-borne GW detector, such as LISA, is a crucial task. It is required to remove the overwhelming laser noise with TDI (a precision of about 30 cm on these LTTs is required to suppress laser noise to required levels), as well as to model the response function of the instrument to GW signals (associated with a much less stringent precision of 300 km).

In this paper, we presented MA-TDIR, an improvement of TDIR, which is an LTT-estimation method based on the minimization of the residual laser noise in the TDI combinations. In LISA, TDIR can be used to calibrate systematic biases of the PRN-based baseline method, cross-check the results, or even be used as a fallback in case of PRN failure. Instead of studying the residual laser noise over the entire frequency band, MA-TDIR modulates the lasers in a small and specific part of the band, increasing artificially the effective laser noise with respect to secondary, unsuppressed noises. As a consequence, MA-TDIR is more robust to signals and data artifacts populating the LISA band and requires shorter integration times to reach the precision required.

We first developed the theoretical framework of MA-TDIR on a two-arm toy model. We derived the likelihood function and used the FIM formalism to evaluate the performance of the method. The formalism was then extended to the full, realistic LISA setup, including the three spacecraft, locked lasers, and a set of three Michelson TDI combinations. The expected precision on the estimated arm lengths was studied when one or two tones are injected in various lasers. We showed that two-tone MA-TDIR LTT estimates are actually competitive with the PRN-based estimates, as they are mostly uncorrelated and have a few meter RMS.

We studied the instrumental feasibility of MA-TDIR and discussed the limiting effect. Modulating tones onto the primary laser has potential to introduce noise in the PDH laser locking to the optical cavity. We showed through an optical test-bed experiment that a laser frequency modulation with a 1 kHz excursion, as considered throughout this manuscript, would not increase the stabilized laser frequency noise above its required level.

Lastly, we ran numerical simulations to check our analytical results. We first checked that our delay optimizer correctly converges to the maximum-likelihood solution in the two-tone case; these two modulating tones, along with laser noise across the LISA band, are reduced to the level of the secondary noises. Then, using an ensemble of 10 000 independent simulations, we verified that the maximum-likelihood delay estimates follow the theoretical distributions derived in the previous sections. In particular, correlations and delay estimate uncertainties are clearly reduced when one uses a pair of modulating tones instead of a single one. Finally, we illustrated with 4 sets of 10 000 simulations the robustness of the proposed MA-TDIR method against in-band data artifacts. We showed that the MA-TDIR LTT uncertainties are unchanged when a large LISA Pathfinder glitch is injected; on the contrary, uncertainties obtained using standard TDIR are largely degraded by the presence of the glitch. Larger and underestimated uncertainties in the LTTs estimates may be such that they do not allow to reach the required laser noise suppression. A solution might be to include TDIR in the global fit, which can easily be done as TDIR is naturally expressed in a Bayesian framework; source parameters would be estimated (to remove loud signals) alongside the LTTs. However, this greatly increases the complexity of the global fit loop, which already presents many challenges [40].

As a conclusion, we remind that the PRN-based method is the current baseline technique for measuring the LTTs and can achieve the precision required by the scientific objectives of the LISA mission. TDIR provides a key redundancy and bias check to the mission-critical measurement of these LTTs. MA-TDIR, the improvement we propose in this paper, only necessitates a modification of the laser locking loop software. As a consequence, we suggest that this possibility is added to the laser system requirements, such that the feasibility of MA-TDIR is not excluded.

## Data availability statement

The data that support the findings of this study are openly available at the following URL/DOI: http://doi.org/10.5281/zenodo.14729644 [29].

## Acknowledgments

The authors thank the LISA Simulation Working Group and the LISA Simulation Expert Group for the lively discussions on all simulation-related activities. They would like to personally thank G Woan and K McKenzie for their insightful feedback. Martin Staab and Gerhard Heinzel acknowledge the support of the German Space Agency, DLR. The work is supported by the Federal Ministry for Economic Affairs and Climate Action based on a decision by the German Bundestag (FKZ 50OQ1801 and FKZ 50OQ2301). This work is also supported by the Max-Planck-Society within the LEGACY ('Low-Frequency Gravitational Wave Astronomy in Space') collaboration (M.IF.A.QOP18098). M S gratefully acknowledges support by the Centre National d'Études Spatiales. Jean-Baptiste Bayle gratefully acknowledges support from UK Space Agency via STFC [ST/W002825/1]. Part of the research was carried out at the Jet Propulsion Laboratory, California Institute of Technology, under a contract with the National Aeronautics and Space Administration (80NM0018D0004).

## Appendix. Expressions for the LISA case

For reproducibility we state the symbolic expressions of the matrix $\Lambda$ used in equation (17). To yield simplified representations we assume equal arms. For the locking configuration N1-L12 it reads

$$\Lambda^{\text{N1-L12}}(f) = \begin{pmatrix} 1 & 0 & -\frac{1-a^2}{2+a} & -\frac{1+a}{2+a} & -\frac{1+a}{2+a} & \frac{a^2}{2+a} \\ 0 & 1 & \frac{1-a^2}{2+a} & -\frac{1+a}{2+a} & \frac{a}{2+a} & -\frac{a(1+a)}{2+a} \\ -\frac{1-a^2}{2+a} & \frac{1-a^2}{2+a} & 1 & \frac{a^2}{2+a} & -\frac{a(1+a)}{2+a} & -\frac{1+a}{2+a} \\ -\frac{1+a}{2+a} & -\frac{1+a}{2+a} & \frac{a^2}{2+a} & 1 & 0 & -\frac{1-a^2}{2+a} \\ -\frac{1+a}{2+a} & \frac{a}{2+a} & -\frac{a(1+a)}{2+a} & 0 & 1 & \frac{1-a^2}{2+a} \\ \frac{a^2}{2+a} & -\frac{a(1+a)}{2+a} & -\frac{1+a}{2+a} & -\frac{1-a^2}{2+a} & \frac{1-a^2}{2+a} & 1 \end{pmatrix}, \quad \text{(A.1)}$$

where we substitute $a = \cos(2\pi f d)$. This matrix is degenerate (its determinant is vanishing), which explains why the FIM for a single tone is singular in the equal-arms limit.

Further assuming delays of $d = 8\,\text{s}$, evaluation of the integral in equation (17) yields the simple expression

$$\Sigma_d \geqslant \frac{1}{T}\frac{S_N(f)}{S_{\dot{p}}(f)}\begin{pmatrix} 1 & 0 & -2+\sqrt{3} & -1+\frac{1}{\sqrt{3}} & -1+\frac{1}{\sqrt{3}} & -2+\frac{4}{\sqrt{3}} \\ 0 & 1 & 2-\sqrt{3} & -1+\frac{1}{\sqrt{3}} & 1-\frac{2}{\sqrt{3}} & 1-\frac{2}{\sqrt{3}} \\ -2+\sqrt{3} & 2-\sqrt{3} & 1 & -2+\frac{4}{\sqrt{3}} & 1-\frac{2}{\sqrt{3}} & -1+\frac{1}{\sqrt{3}} \\ -1+\frac{1}{\sqrt{3}} & -1+\frac{1}{\sqrt{3}} & -2+\frac{4}{\sqrt{3}} & 1 & 0 & -2+\sqrt{3} \\ -1+\frac{1}{\sqrt{3}} & 1-\frac{2}{\sqrt{3}} & 1-\frac{2}{\sqrt{3}} & 0 & 1 & 2-\sqrt{3} \\ -2+\frac{4}{\sqrt{3}} & 1-\frac{2}{\sqrt{3}} & -1+\frac{1}{\sqrt{3}} & -2+\sqrt{3} & 2-\sqrt{3} & 1 \end{pmatrix}^{-1}. \tag{A.2}$$

## ORCID iDs

Jean-Baptiste Bayle ● 0000-0001-7629-6555
Martin Staab ● 0000-0001-5036-6586
Emily Rose Rees ● 0000-0003-0112-6716
Robert Spero ● 0000-0002-8240-3369
Gerhard Heinzel ● 0000-0003-1661-7868